\documentclass{jaa}
\usepackage{natbib}
\usepackage{hyperref}
\hypersetup{
    colorlinks=true,
    linkcolor=blue,
    filecolor=magenta,      
    urlcolor=cyan,
    pdftitle={Overleaf Example},
    citecolor=red,
    }
\usepackage{graphicx}

\begin{document}\sloppy

%%paper title
%%For line breaks \\ can be used within title
\title{Innovative Web Tool for Remote Data Acquisition and Analysis: \\ Customized for SKA Low frequency Beamforming Test Bed LPDA Array at Gauribidanur Radio Observatory}

%%author names are separated by comma (,)
%%use \and before the last author name
%%use a * along with the number separated by comma
%% for the  author for correspondence
%%\textsuperscript{number} is used for affiliation
%%\affilOne, \affilTwo etc., upto \affilTwentyfive is possible
%%Please note the first letter after \affil is capitalised in the command
%%

\author{Anumanchi Agastya Sai Ram Likhit\textsuperscript{1,2}, Katta Naveen \textsuperscript{1,2}, Arul Pandian B\textsuperscript{1}, Abhishek R \textsuperscript{1}, Prabu T\textsuperscript{1}}
\affilTwo{\textsuperscript{2}  Department of Physics, Indian Institute of Science Education and Research Bhopal, 462066, India}
\affilOne{\textsuperscript{1} Electronics Engineering Group, Raman Research Institute, 560080, India \\}

%%escape two column mode for title, affiliation and abstract
%%by giving \twocolumn command as shown

\twocolumn[{

\maketitle

%%include \corres to print the corresponding author Email id
\corres{astropi.2003@gmail.com}

%%include \msinfo for
%%manuscript information such as
%%received, revised and accepted dates
%%
%\msinfo{1 January 2015}{1 January 2015}

%%abstract
\begin{abstract}
With the evolution of radio astronomy, related education and training, the demand for scalable, efficient, and remote systems in data acquisition, storage, and analysis has significantly increased. Addressing this need, we have developed a web interface for a log-periodic dipole antenna (LPDA) array integral to the SKA Test activities at the Gauribidanur Radio Observatory (77.428 E, 13.603 N). This interface, employing Python-based technologies such as Streamlit and PyVISA, along with SCPI commands, offers a seamless and user-friendly experience. Our solution introduces a unique data acquisition approach, employing SCPI through Python to communicate with the setup's Data Acquisition System (DAS). The web interface, accessible remotely via a secure WLAN network or VPN, facilitates user-initiated observations and comprehensive logging and offers advanced features like manual RFI masking, transit plotting, and fringe plot analysis. Additionally, it acts as a data hub, allowing for the remote downloading of observational data. These capabilities significantly enhance the user's ability to conduct detailed post-observation data analysis. The effectiveness of this interface is further demonstrated through a successful solar transit observation, validating its utility and accuracy in real-world astronomical applications. The applications of this web tool are expandable and can be tailored according to the Observatory's Goals and Instrumentation as well as for the growing radio astronomy instrumentation and observing facilities coming up at various educational institutions.
\end{abstract}

%%insert keywords separated by 3 hyphens using \keywords{words}

}]
%%close the twocolumn escape here

%%include \doinum{number}for the DOI number in the header
%%include \volnum{number} for the volume number in the header
%%include \year{yyyy} for  year of publication in the header
%%include \pgrange{num--num} page range of article in the header
%%include \artcitid{num} for the article citation id
%%include \lp to print last page of the article
%%include \setcounter{page}{pagenum} for the exact starting page of the article

\doinum{12.3456/s78910-011-012-3}
\artcitid{\#\#\#\#}
\volnum{000}
\year{0000}
\pgrange{1--}
\setcounter{page}{1}
\lp{1}

\section{Introduction}
The field of radio astronomy has witnessed a rapid evolution, driven by the quest to explore the universe more deeply. We also observe the increased awareness and capabilities to set up small-scale radio telescope observational facilities that are coming up at the diverse educational institutes specialising in astronomy. This evolution brings with it a demand for advanced observational capabilities, particularly in terms of data acquisition, storage, and analysis. A significant challenge in this domain is the development of systems that are not only efficient and scalable but also capable of remote operation, given the often inaccessible locations of radio observatories.

These expanding needs, particularly at the Gauribidanur Radio Observatory, a hub of important astronomical research, are what drive our work. A sophisticated approach is required to conduct observations and handle the data remotely using the observatory's log-periodic dipole antenna (LPDA) array, which is used for Square Kilometre Array (SKA) India \citep{Gupta2023-tr} test activities. Traditional methods often fall short in terms of efficiency, user-friendliness, and flexibility, especially when it comes to remote accessibility.

Recognizing these challenges, our goal was to create a solution that not only addresses these technical demands but also enhances the overall user experience for astronomers and researchers. A significant response to this need emerged with the development of a web interface that utilizes state-of-the-art Python-based technologies like Streamlit and PyVISA, along with Standard Commands for Programmable Instruments (SCPI). The design of this interface streamlines the process of data acquisition and offers robust features for manual Radio Frequency Interference (RFI) detection and masking and transit plotting—all accessible remotely, a crucial feature aiding modern astronomical research.

The design presented in this paper has been developed and successfully operational in the observatory for the past six months for a new broadband two-element radio interferometer. The hardware details of the interferometer implementation are discussed in detail in \citep{Dora,Abhishek}, and this paper presents the salient aspects of the novel web-based remote data acquisition and data analysis features developed for the interferometer. 

%The web tool we are proposing is one-of-a-kind and unique, as there are no other existing web tools that offer the same comprehensive functionality as ours. While there are various tools and systems used in radio astronomy for data acquisition and analysis like \citep{Nagios, Sashikant} which were used in the MWA telescope, which is a precursor telescope of the SKA.  , none integrate the full range of capabilities—including remote data acquisition, RFI masking, and advanced data processing—into a single, user-friendly interface accessible over a secure network. This unique combination of features positions our tool as a unique one, addressing specific needs in radio astronomy, particularly for observatories with remote operation requirements, such as the Gauribidanur Radio Observatory.

Astronomy communities have developed various software tools for monitoring and operating radio telescopes. We describe two SKA precursor MWA telescope tools: the Common Gateway Interface (CGI-BIN) based wMARC \citep{Sashikant} receiver commissioning and NAGIOS-based telescope monitoring tools. The receiver commissioning required executing several discretely developed standalone codes and OS scripts in a flexible sequence. The wMARC integrated these executables to provide a customizable menu-driven environment, and a remote client's web browser could invoke the executables using HTTP and HTTPS calls. The tool was used extensively during the MWA telescope's commissioning phase. The Nagios \citep{Nagios} is an event monitoring system typically valuable for monitoring servers, network switches and applications. A Nagios-based system was used to monitor the status of the MWA telescope functions. Inherent to the Nagios is extensive logging of the events, system configuration retrieval, and alert mechanisms with an interface that includes email message alerts and authorization mechanisms. This system uses cgi-bin and plugins. Plugins are external executables written in any language. Sometimes,  a CGI-based implementation concerns sensitive networks, exposing the system to external programs. 

 The tool presented in this work provides comprehensive functionalities for remotely operating a simple radio interferometric telescope and collecting and analyzing both current and archived data. It is based on a free and open-source framework, namely the Streamlit, which is popular among machine learning and data science web applications.

The paper's outline is as follows: Section 2 presents the background of our work, focusing on the log-periodic dipole array (LPDA) we employed. Section 3 elaborates on the signal flow within our observatory's model telescope setup. Section 4 discusses the data acquisition methods discovered, detailing the chosen observation approach. Section 5 covers the data processing pipeline for the collected data. The design and implementation of our web interface are outlined in Section 6, with Section 7 explaining its functionalities. Section 8 addresses testing and validation of the web tool, including its application in actual astronomical observations. We conclude by reflecting on the web tool's success and its potential for broader application.

\section{The Log Periodic Dipole Antenna Array}
\begin{figure}[h]
    \centering
    \includegraphics[width=3in]{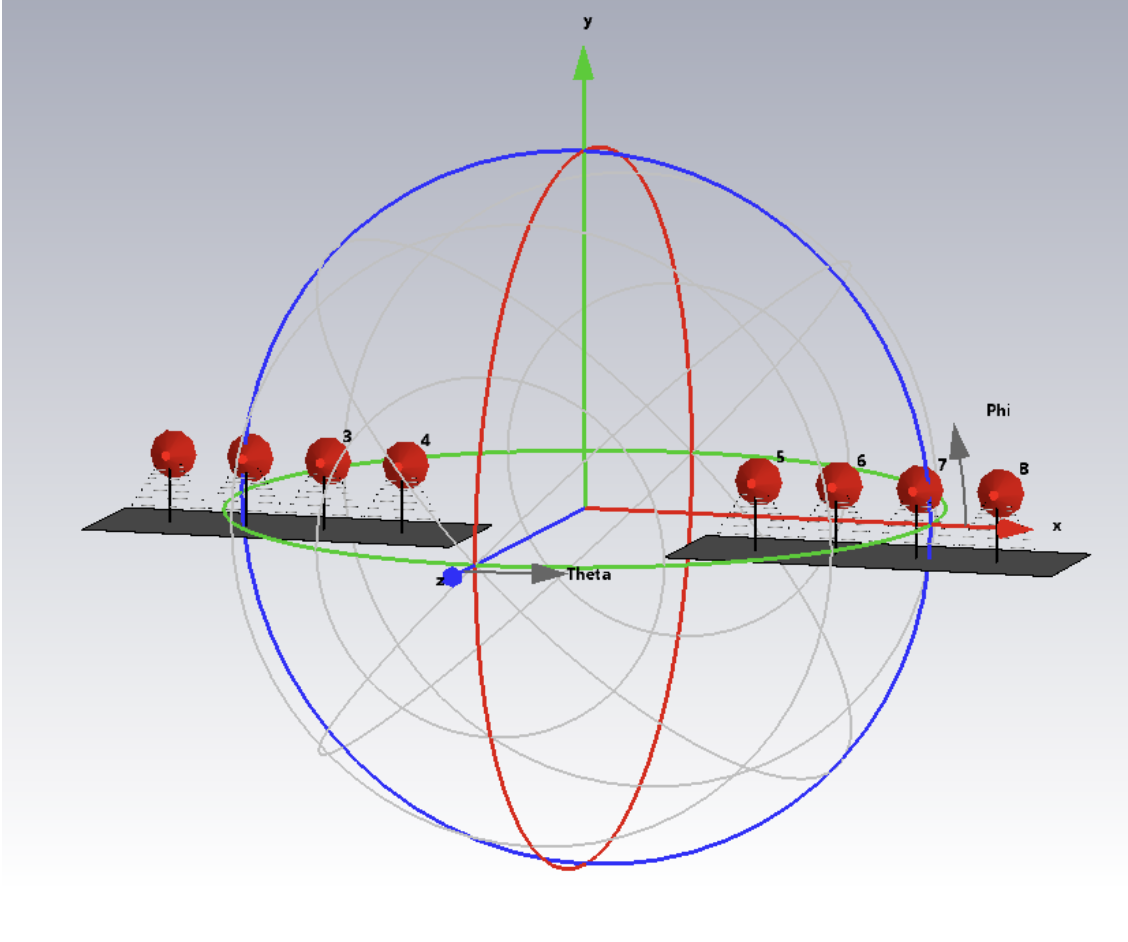}
    \caption{\centering LPDA Array at Gauribidanur Radio Observatory}
    \label{fig:1}
\end{figure}

At the Gauribidanur Radio Observatory, we have a set-up of eight log-periodic dipole antennas (These are similar to what was mentioned in \cite{Raghunathan2023-mc}. Which are organized into two elements, each containing four antennas. As shown in the image in Fig.\ref{fig:1}, the antennas are aligned in an east-west direction and operate within the 150–350 MHz range,overlapping with the upcoming SKA Low radio telescope.

\begin{figure}[h!]
    \centering
    \includegraphics[width=3in]{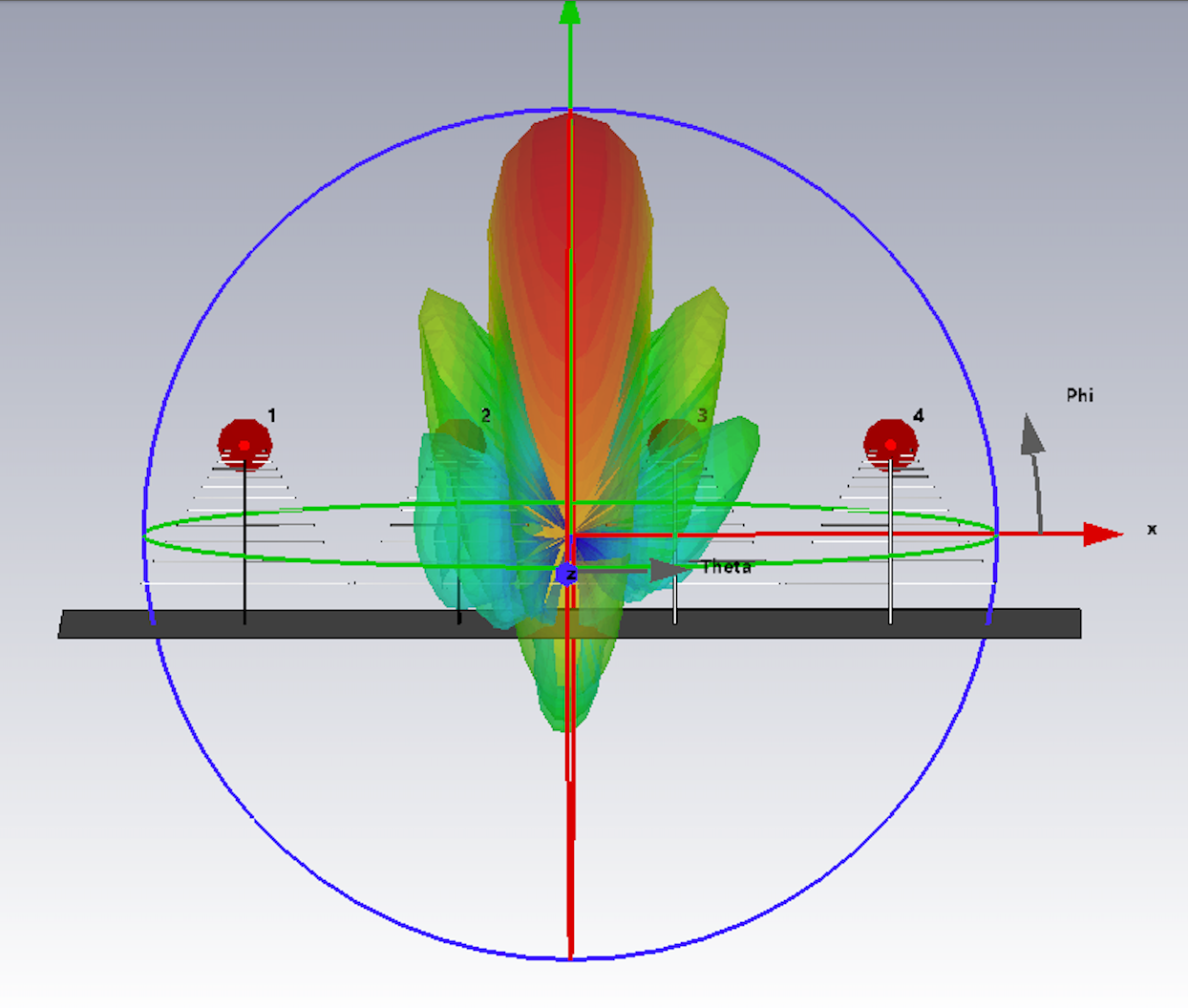}
    \caption{\centering Simulated 3D Far-Field Radiation Pattern of a Single Element at 200MHz}
    \label{fig:2}
\end{figure}

The baseline of the two element antenna setup spread over 13.1 m along East-West (consisting of about 5 m and 5.3 m for two elements, with a free space of about  2.8  m between the elements). Using the actual parameters of the setup, we simulate each element in an antenna simulation software CST. The simulation shows, at a frequency of 200 MHz, the main lobe is directed towards the zenith with a magnitude of 11.6 dBi and an angular width of 18.1 degrees at the 3 dB point. The side lobe measured level is -12.7 dB, which indicates effective sidelobe suppression as it is sufficiently below the main lobe peak. The two element beam pattern obtained from the simulation is shown in  Fig.\ref{fig:2}.

These characteristics are crucial for the accurate interpretation of the observed radio signals and for ensuring that the primary reception is from the intended direction of a celestial source.
Our objective is to create a web tool that can efficiently manage all the observations conducted on this log periodic dipole antenna array. This tool includes tasks such as data collection, storage, and processing. Before exploring the web tool, let us review the RF path of the setup.

\section{RF Signal Path for the Log Periodic Dipole Antenna Array}
\label{rfpath}
The signal processing chain for each element of the Log-Periodic Dipole Array (LPDA) at the Gauribidanur Radio Observatory is designed in the following way: This RF path begins with the collection of radio signals by four LPDA antennas, which are engineered to operate over a broad frequency range. These antennas are particularly sensitive to the target frequency band of 150 MHz to 350 MHz, making them ideal for the observatory's research purposes.

\begin{figure}[h]
    \centering
    \includegraphics[width=3in]{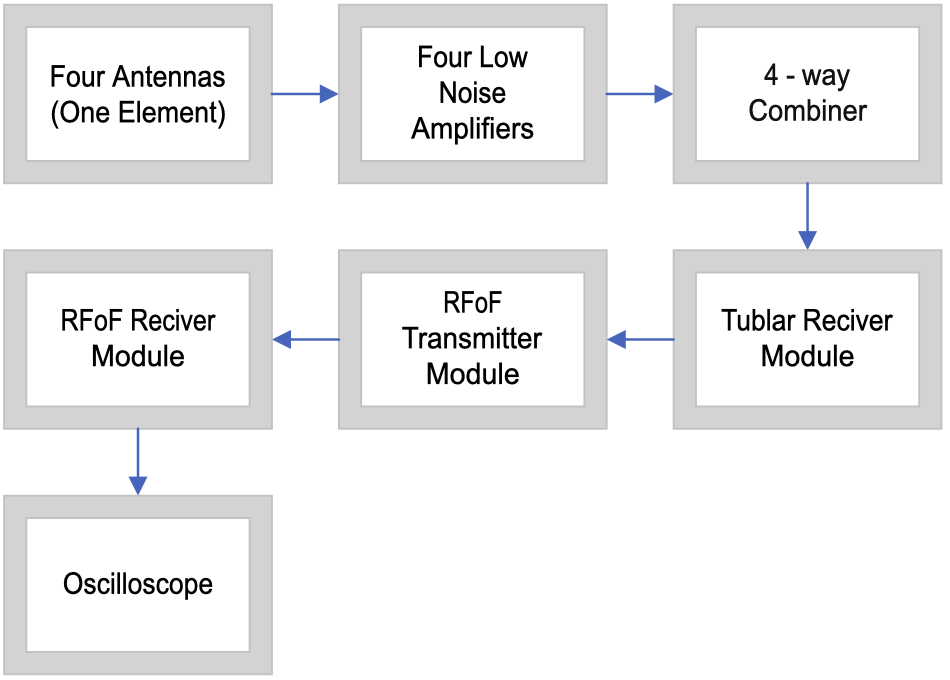}
    \caption{\centering Radio frequency signal path tailored for an element}
    \label{fig:3}
\end{figure}

Signals from the celestial radio sources are very weak, so we use the above-mentioned sensitive RF path (Fig \ref{fig:3}) to preserve the signal strength for the analysis. Upon capture by the antennas, the signals are individually amplified by custom-designed low-noise amplifiers (LNA). These LNAs, custom developed in-house \cite{kbr}, are critical in boosting the signal strength while minimizing the system noise, thereby preserving the quality of the received astronomical signals. Following initial amplification, the signals from the four antennas are combined using a 4-way combiner. This process not only increases the overall signal strength but also enhances the signal-to-noise ratio.

The combined signal is then processed through a front-end receiver module, another custom component designed at RRI, containing a series of bandpass filters, high-pass filters, amplifiers, and attenuators. These components collaboratively filter out frequencies outside the intended band, provide further amplification, and adjust the signal level, ensuring the output signal is precisely within the 150 MHz to 350 MHz band with a gain of +53 dB. Next, the signal is sent to an RFoF (Radio Frequency over Fiber) transmitter module, which converts the RF signal into an optical signal to be transmitted over a fiber-optic cable. This method of transmission preserves the quality of the signal over long distances with minimal loss. The corresponding RFoF receiver module then converts the optical signal back into an RF signal for electronic processing.

Finally, the RF signal is fed into an oscilloscope, a device that visualizes the signal in the time domain. This allows for real-time observation and analysis of the signal's properties, such as amplitude and frequency. The oscilloscope can be utilized to confirm the functionality of the complete RF channel by examining the signal properties and confirming they correspond to the anticipated 150 MHz to 350 MHz frequency range and the overall amplification of 63 dB.

Thus the RF path is carefully engineered to capture, amplify, filter, and analyze radio signals within a specific frequency range, enabling astronomical observations in the SKA Low range at the Gauribidanur radio observatory using the Log periodic dipole antenna array.

\section{Data Acquisitions Methods for Log Periodic Dipole Antenna Array}
Accurate data gathering and control are essential for observing transient phenomena, continuum source transits and for tracking them. The Tektronix Mixed Signal Oscilloscope MSO-3054\footnote[1]{More about Tektronix Mixed Signal Oscilloscope MSO-3054 in the Appendix A}, shown in Figure \ref{fig:oscilloscope}, served as a crucial data gathering device for our observation.

\begin{figure}[h]
    \centering
    \includegraphics[width=3in]{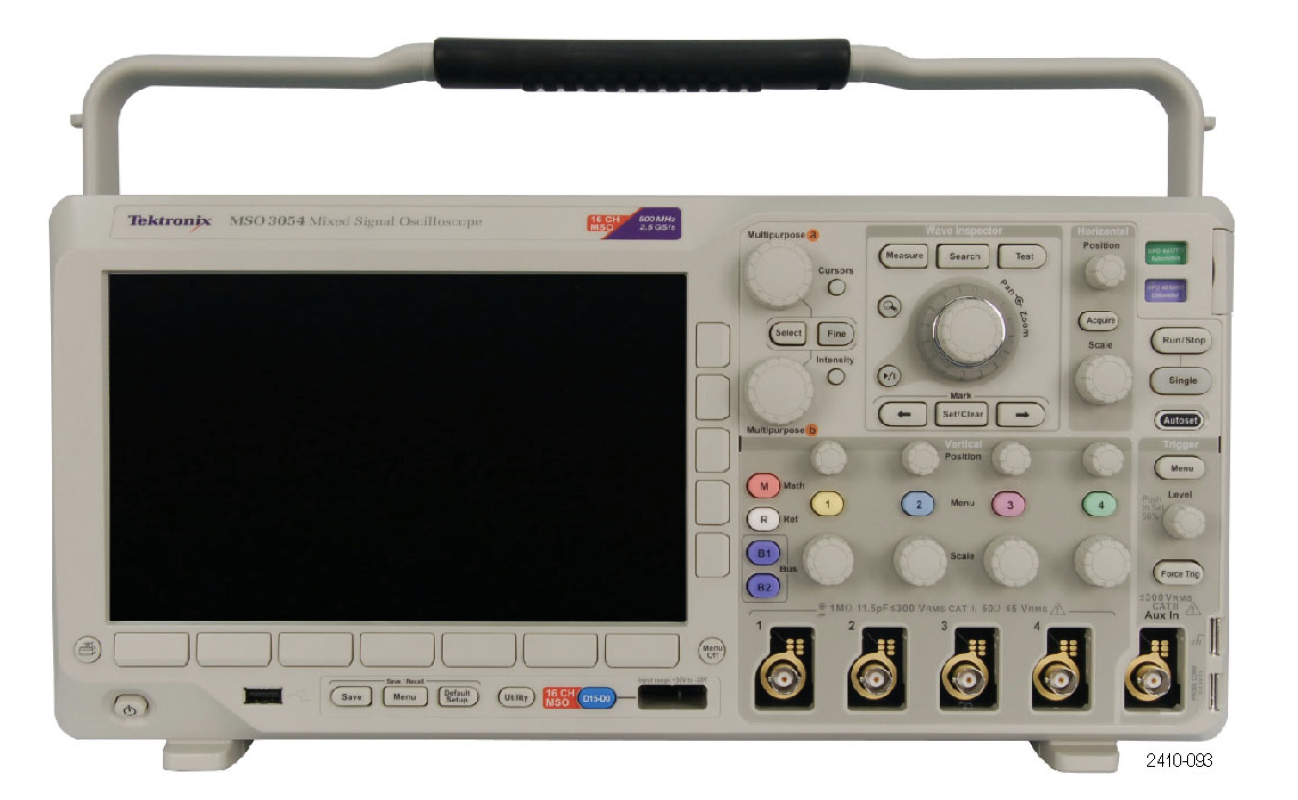}
    \caption{\centering Tektronix MSO3054 Mixed Signal Oscilloscope- Backend for our Observstions}
    \label{fig:oscilloscope}
\end{figure}

The RF signals from each antenna element are fed into two radio frequency (RF) input channels of this oscilloscope. The oscilloscope is connected to the local network via Ethernet, enabling remote access. In the following subsections, we present two innovative methods for data acquisition that we utilized in our work.

 \textbf{1. Automated Remote Control with e* Scope : } One of the primary challenges in our data acquisition was obtaining dual-channel data synchronously to facilitate correlation and phased array mode of observations. To address this, a method was developed using the oscilloscope's built-in remote viewing and control feature, e*Scope.

 \begin{figure}[h]
    \centering
    \includegraphics[width=3in]{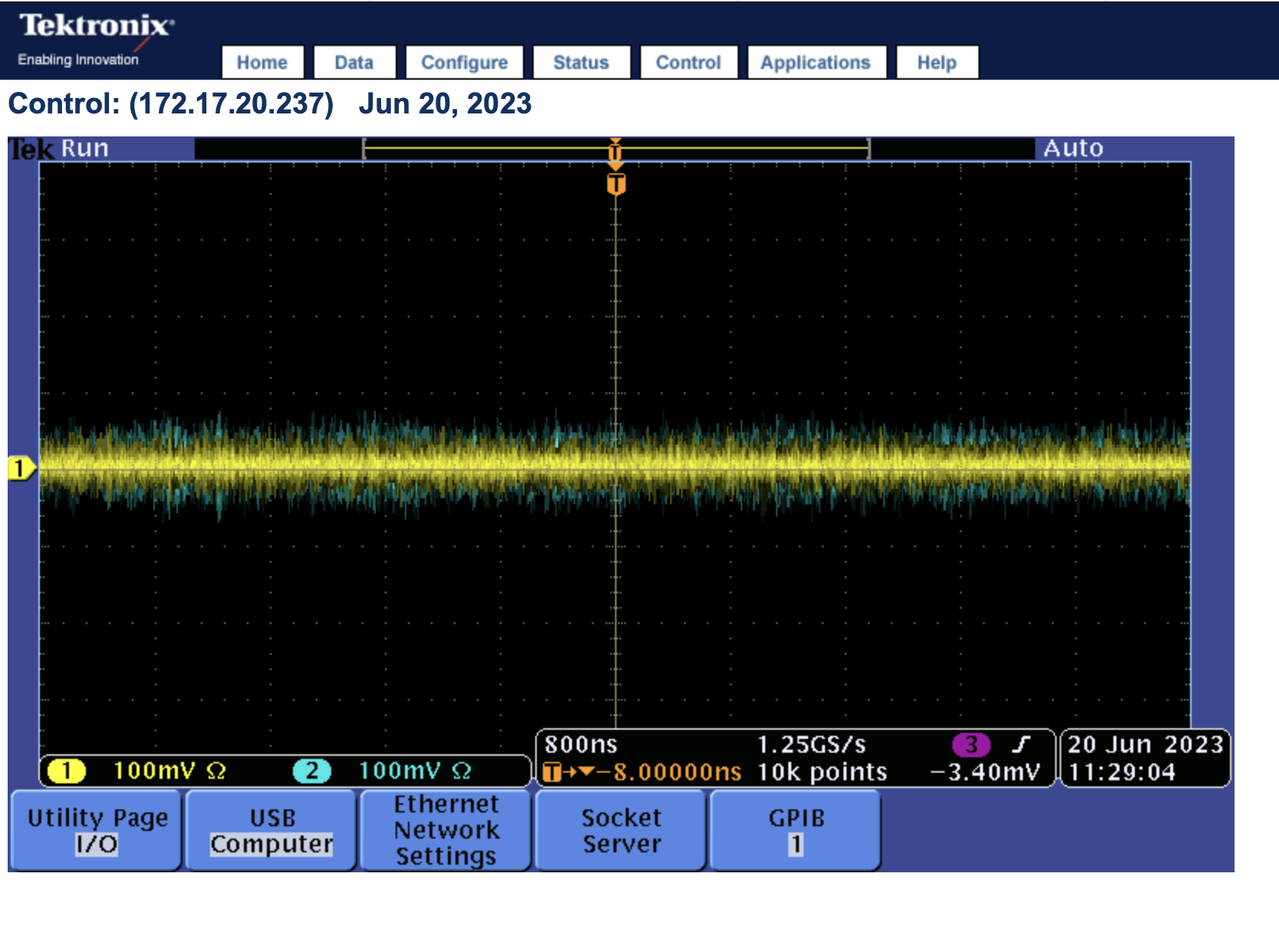}
    \caption{\centering e*Scope feature of the oscilloscope}
    \label{fig:oscilloscope}
\end{figure}

As mentioned earlier, an Ethernet cable connects the oscilloscope to the internet. The web interface of the oscilloscope allows access for remote control of the data. This interface provides an option to save waveform as a comma-separated-value (CSV) data format files containing data for all channels synchronously gathered at the same time. To achieve a continual data acquisition over extended periods, such as full-day observations, we wrote a Python program utilizing the Selenium python package \citep{selinium,Gojare} , which provided a robust connection to the oscilloscope's web interface. This package helps to integrate the oscilloscope's front panel operations into a python program scripts. By analyzing the webpage's structure, we discover the URL(universal resource locator) for the "Save Waveform" button and automate the task of clicking it at regular intervals. The user could set the start time, end time, and interval, and the program would automatically open the URL, initiating the save waveform action accordingly. The oscilloscope saved the gathered two channel data as a CSV format data files in to a USB storage device. As the data gathering continues, the oscilloscope saves each scan data in a separate file along with the  time information. The oscilloscope named the files sequentially, such as \(tek\_ALL\_0000.csv\), \(tek\_ALL\_0001.csv\), and so on. The oscilloscope collects the data without any manual intervention, ensuring accurate timing for repeated scans.

While this method allowed for precise data collection at simultaneous intervals for two channels, it did come with specific limitations. Firstly, the USB storage capacity could become a constraint depending on the size of the data files and the duration of the observation. Secondly, the oscilloscope can only save up to 9999 CSV files, after which it will start rewriting the first file. This means careful planning is required for very long observations.

\textbf{2. Direct Data Transfer Using PyVISA: } The second method of data acquisition was designed to overcome the limitations of the first, providing a more flexible and robust solution, particularly for longer observations. This approach utilized the run/stop command of the oscilloscope, offering a novel way to collect dual-channel data synchronously at the same time.

By using the Run/Stop button, the oscilloscope is made to freeze by holding the gathered signal waveform data at a particular time t1, effectively stopping the live visualization and subsequent acquisition of the signals until a run command is used. Once the signal gathering is stopped, the waveform data of each channel is transferred to the computer using PyVISA \citep{Grecco2023}, a Python-based instrument control package. After data transfer, the live signal is promptly resumed by pressing the run/stop button, then pausing again after a period of t2 is repeated throughout the observation. This data transfer based on PyVISA appeared to be faster.

This repeated run/stop requires certain minimum time gaps to be adhered to to maintain data integrity. The time constraints come from the following elements: a) time to gather one record signal waveform, typically a few thousand to a million times the chosen sampling rate in the oscilloscope, b) time to transfer the recorded data over the internet as standard TCP packets, c) the oscilloscope's internal turnaround time to restart a new acquisition.

\section{Data Processing}
The data gathering pipeline collects the data at regular intervals over the selected period and saves the data in CSV format file for each acquisition data-packet that provides voltage readings. Each data packet containing 10,000 measurement points or voltage readings. We organized these packets into arrays for each channel, creating extensive datasets ready for analysis. We divided these packets into smaller segments, each with 1,024 data points, resulting in nine segments per packet.

To convert this data from the time domain to the frequency domain, we applied the Real Fast Fourier Transform (RFFT). The 'Real' in RFFT indicates the inputs are only real numbers, optimizing the computation. This method converts the time-domain signal into its constituent frequencies, producing 513 output points per segment. We discarded the first point to eliminate the DC voltage component, leaving 512 frequency bins. Each bin represents a bandwidth of 1.22 MHz, calculated by dividing the oscilloscope's 0 to 625 MHz range by 512.

We then calculated the signal's power by squaring each RFFT point's magnitude. By averaging these squared values, we obtained a 512-point array for each acquisition, which measures signal power. To express this power in dBm, a standard unit for spectral density, we took the -10 logarithm of these values. Averaging the power across the same frequency bins produced a spectrum, and averaging across all bins for each packet gave us a time series of signal power. 

\subsection{Correlation Algorithm}

The correlation analyzes the relationship between signals captured by two different channels (CH1 and CH2) from the antenna array. We divide the data from each channel into smaller segments to facilitate detailed analysis. We specifically break down each packet of data, containing 10,000 voltage readings, into segments of 1,024 points. For each segment, we apply the Real Fast Fourier Transform (RFFT) to convert the time-domain signal into the frequency domain. This step is crucial because it allows us to analyze the frequency components of the signal. In this process, we discard the first frequency channel of the RFFT output to eliminate the DC component, focusing on the remaining frequency data.

After computing the RFFT for each segment, we calculate the power by taking the square of the magnitude of the Fourier-transformed data. We then average this power across all segments within each packet to obtain a representative power spectrum for each packet.

To measure the similarity between the signals from the two channels, we compute the cross-correlation of the Fourier-transformed data. This involves multiplying each segment's Fourier-transformed data from CH1 by the complex conjugate of the corresponding segment from CH2. The resulting correlated data provides a measure for the  the similarities between the two signals, allowing us to identify any time delays or phase differences. Finally, we calculate the average of the correlated data's absolute values for each packet. This mean value gives a summary of the correlation between CH1 and CH2, which is critical for understanding the relative positioning and timing of the observed signals.

\subsection{RFI Masking }
  
Radio Frequency Interference (RFI) is the unwanted disturbance in the observed radio signal that is caused by man-made sources like satellites and local electronic equipment. These interferences can corrupt the data collected by the telescope, making it challenging to extract meaningful information. Effective RFI mitigation is crucial in radio astronomy to ensure that the data accurately reflects the true signals from the sky, free from unwanted noise sources.

To address RFI in our data processing pipeline, we focused on analyzing the power of each frequency bin over time.  Specifically, heat maps are plotted to represent the power levels across different frequency bins throughout the observation period. These heat maps allowed us to visually identify the bins and corresponding time intervals where RFI was present. Once we identified these regions, we manually mask the affected bins by setting their values to zero.

For instance, we observed RFI from satellites as well as interference from the local oscillator, which introduced a consistent, predictable noise in certain bins. By manually identifying and masking these interferences, we were able to reduce their impact on the data. However, while this manual masking process is effective, it is labor-intensive and leaves room for improvement. There is significant potential for automating this process using machine learning techniques \citep{Mosiane} , which could identify and mitigate RFI more efficiently.

\section{Design and Implementation of Web Tool}
To address the challenges of remote data acquisition and analysis, we developed a novel Web Tool that allows simultaneous data acquisition from the two elements at regular intervals and is remotely accessible from the Raman Research Institute campus or also can also be accessed globally via a Virtual Private Network (VPN). The system offers a range of functions, including data downloading, logging of observations and instrument settings, data processing and Radio Frequency Interference (RFI) masking. All these features are seamlessly integrated into a user-friendly web interface, providing a comprehensive solution for data management needed in the radio telescope observations.

\begin{figure}[h]
    \centering
    \includegraphics[width=3.5in]{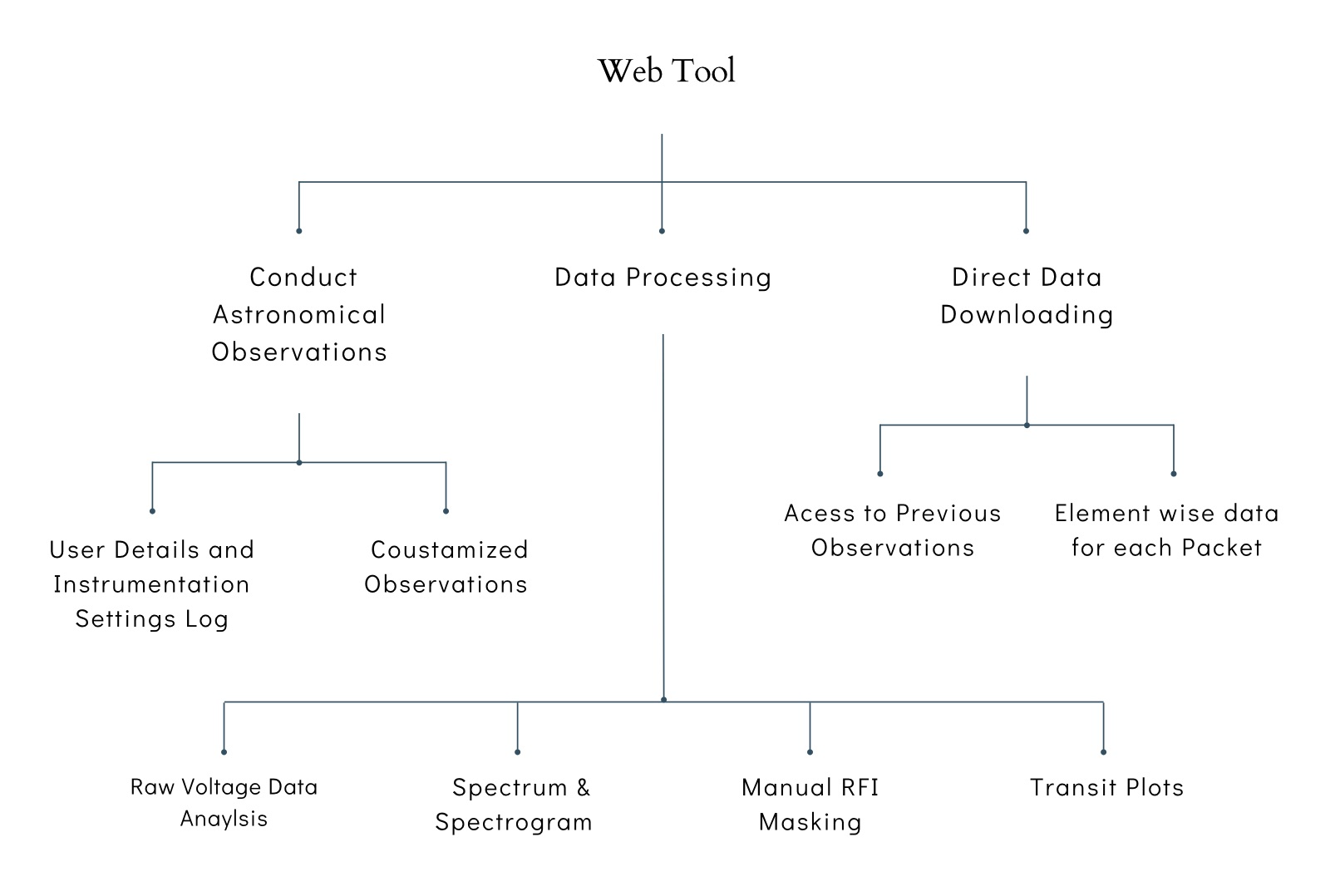}
    \caption{\centering Outline of the web tool functionalities.}
    \label{fig:functionality}
\end{figure}
\textit{Streamlit} \citep{streamlit} is an open-source Python library that simplifies creating web applications for data science and machine learning. With its user-friendly design, Streamlit allows to transform data scripts into shareable web apps. Given these features, we chose Streamlit for designing a web tool that serves as a hub for observations, data storage, and data analysis. The platform offers numerous advantages, such as  rapid prototyping, interactive analysis, extensive customization options and ease of use. making it an ideal choice for our project.\\
We developed the Python-based web tool using Streamlit, hosted on a local system at the Gauribidanur Radio Observatory and connected to remote users in the main campus via WLAN.

\begin{figure}[h]
    \centering
    \includegraphics[width=3.5in]{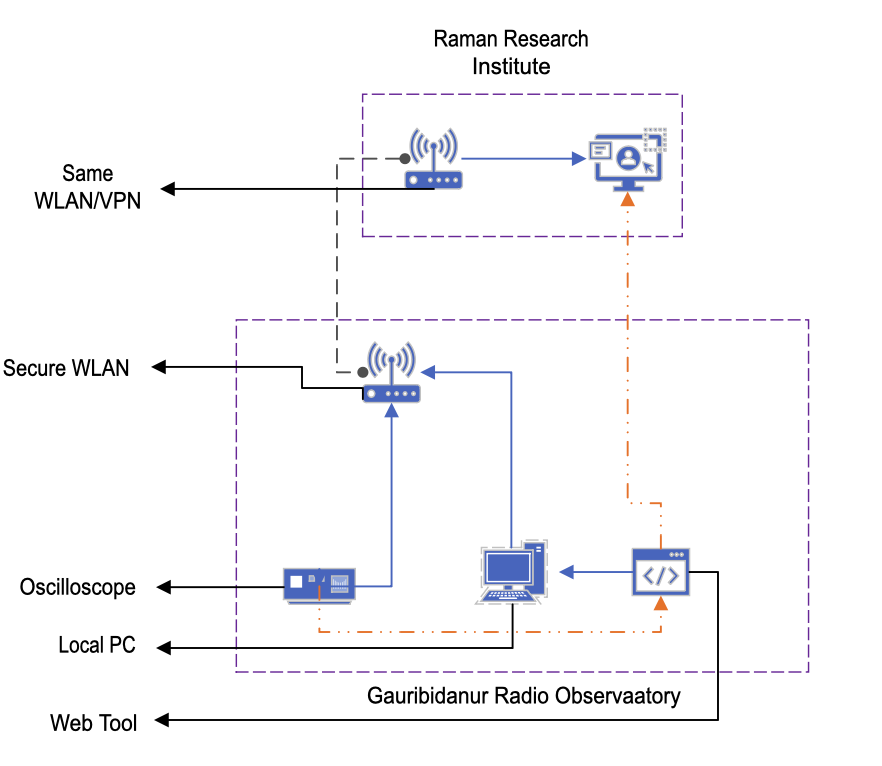}
    \caption{Schematic representation of the network configuration}
    \label{fig:fw}
\end{figure}

A Schematic representation of the network configuration between the Gauribidanur Radio Observatory and the Raman Research Institute in the Fig \ref{fig:fw}. The diagram illustrates the data flow from the observatory's oscilloscope to the web tool hosted on a local PC, accessible via a secure WLAN or remotely through a VPN.

\section{Functionalities of Web Tool}

The web tool offers a broad array of functionalities, ranging from initiating observations to conducting advanced data analysis and plotting results. The main page of the web interface displays an image of the log-periodic dipole antenna array, welcoming users and inviting them to navigate through three distinct sections: (i) Data Acquisition, (ii) Data Center, and (iii) Data Processing. The following text delve into each of these sections of the web tool to explore what they offer to the user.

\begin{figure*}[h]
    \centering
    \includegraphics[width=5.4in]{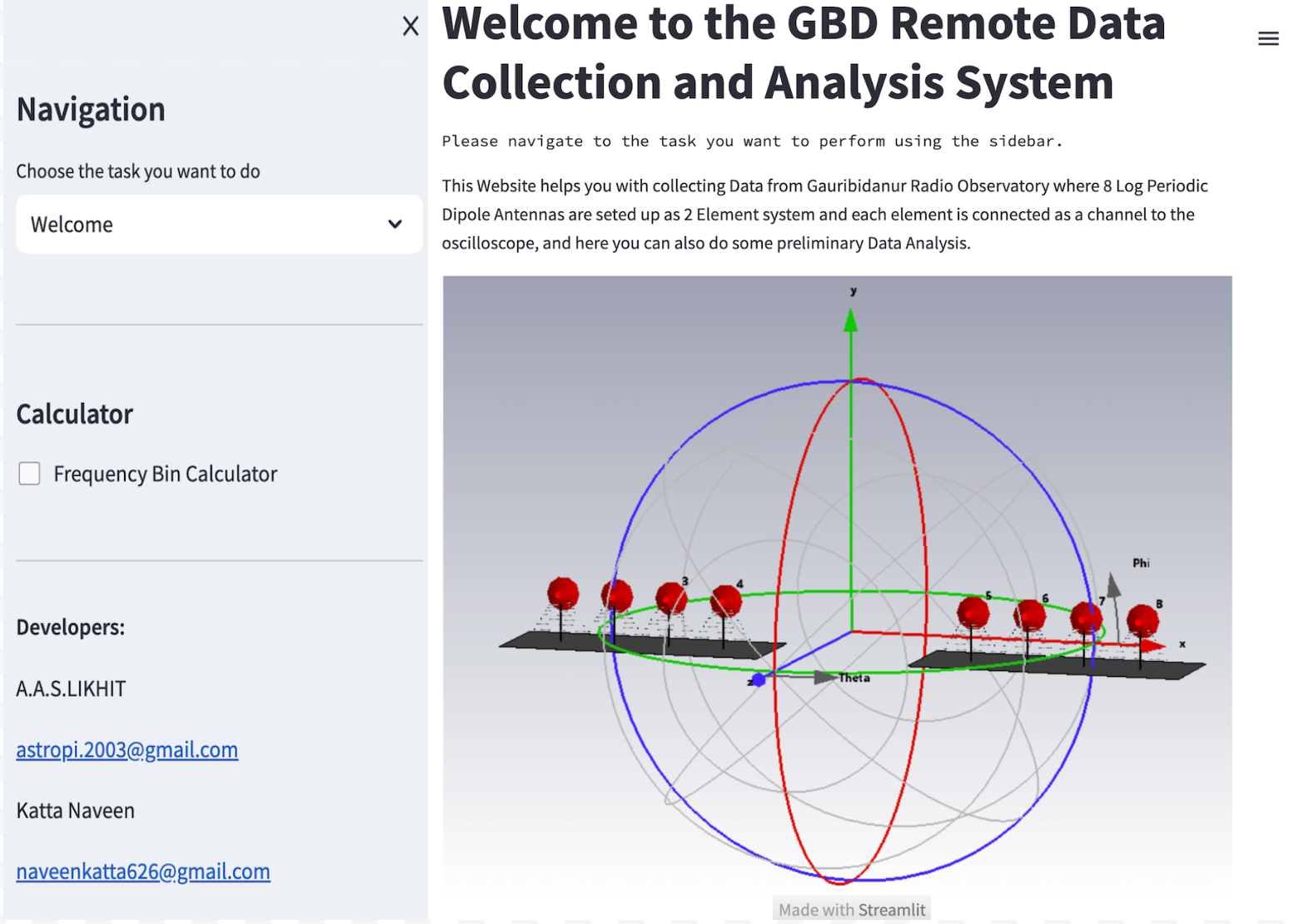}
    \caption{Home Page of the Gauribidanur radio observatory LPDA Observation Tool (GLOT), displaying the Navigation Bar with available features and the interactive 3D visualization of the LPDA array setup.}
    \label{fig:intro}
\end{figure*}

\subsection{Data Acquisition}

This section of the web tool enables users to initiate observations with the antenna array. Users can manage various aspects of the observation, such as setting the start and end times of the observation, selecting the number of data channels to collect (in our case, we have two elements), and adjusting the time delay between acquisitions based on the specific requirements of their observations. Also, we have to enter a name for the observation. Using the name, it creates a folder for saving the collected data.

\begin{figure}[h]
    \centering
    \includegraphics[width=3.4in]{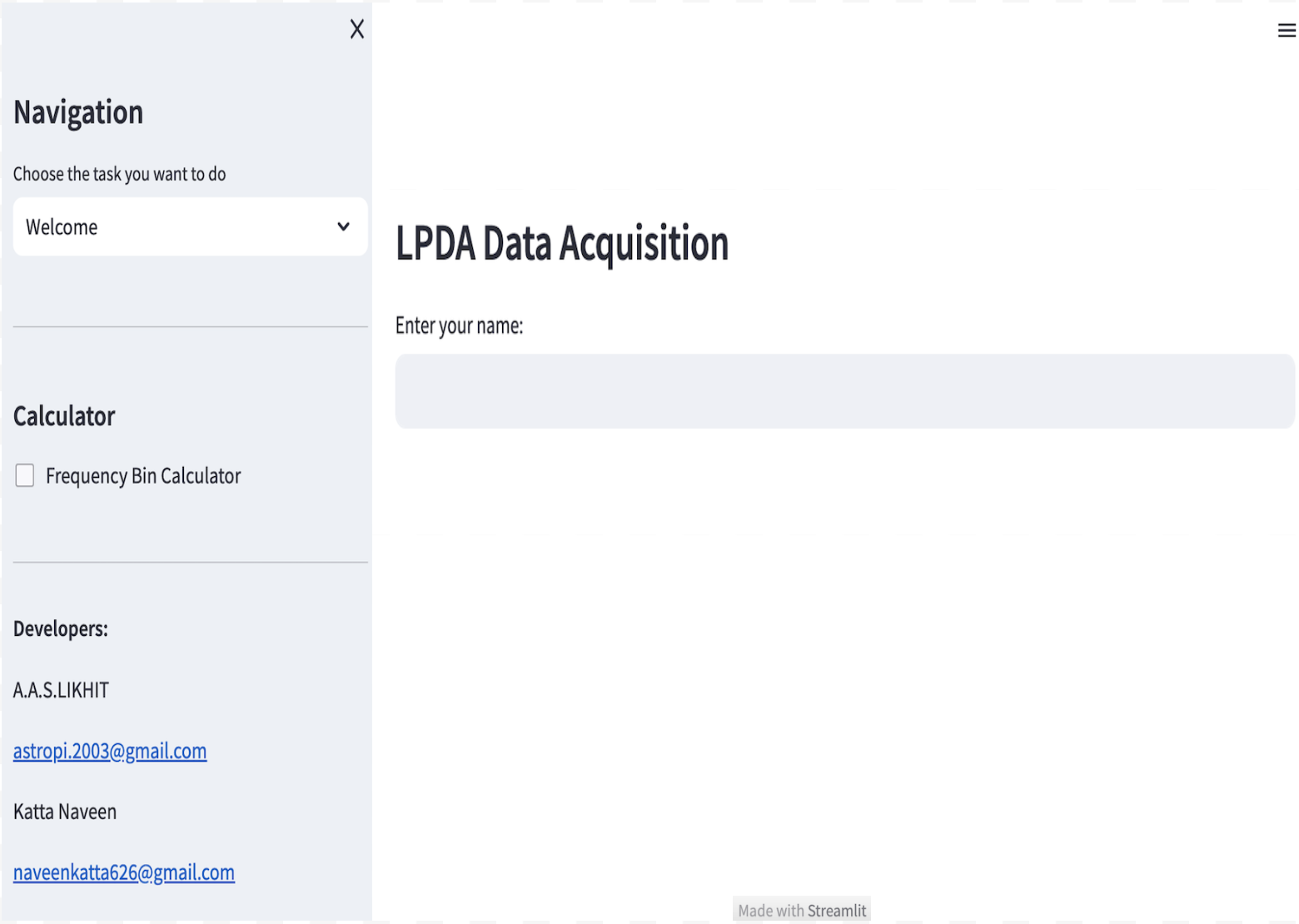}
    \caption{Initial user authentication prompt asking user for their name to proceed further to start an observation.}
    \label{fig:da1}
\end{figure}

\begin{figure}[h]
    \centering
    \includegraphics[width=3.4in]{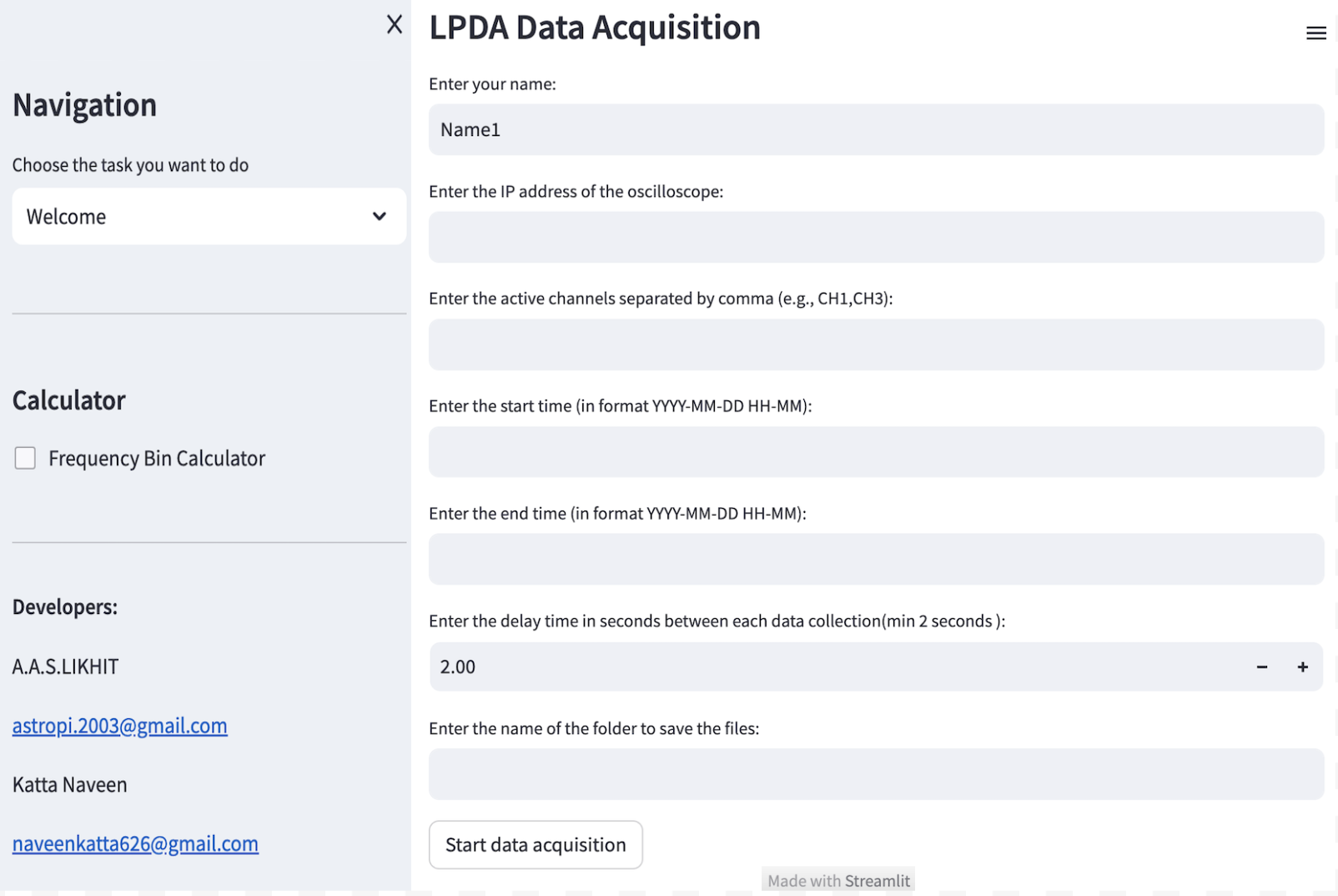}
    \caption{Data Acquisition section of the web page with prompts for user inputs for the observation settings needed to start the data acquisition.}
    \label{fig:da2}
\end{figure}

Upon entering this page, the user is prompted to input their username, Fig \ref{fig:da1} illustrates the user interface requesting the username entry. After successful verification, options to start the observation become available as shown in Fig \ref{fig:da2}. The web tool is designed to track and log all inputs provided by the user into a log file, continuously appending to a CSV file. Furthermore, it communicates with the oscilloscope, generating a new text file with detailed oscilloscope settings at the start of each observation. These files are distinctly labeled with the date and time of the observation and are stored on the local system.

Access to these files is restricted to administrators, allowing them to review previous observation details. This level of inspection into user inputs and oscilloscope settings provides comprehensive information about the observations conducted.
The various features incorporated in the Web-Tool allow the users to customize the observation needs.
Currently, the Web-Tool has been regularly used for periodic remote observations, for a long time.

\subsection{Data Center}
This section of the web tool allows users to log and review previous observations Data. Users can view  individual files for each observation.

Additionally, as depicted in Figure \ref{fig:datacenter}, there are options to download individual files or to download all files from an observation in a zipped folder. This functionality enables users to exclusively download data collected from the antenna array at gauribidanur radio observatory and conduct their own analysis.

\begin{figure}[h]
    \centering
    \includegraphics[width=3.4in]{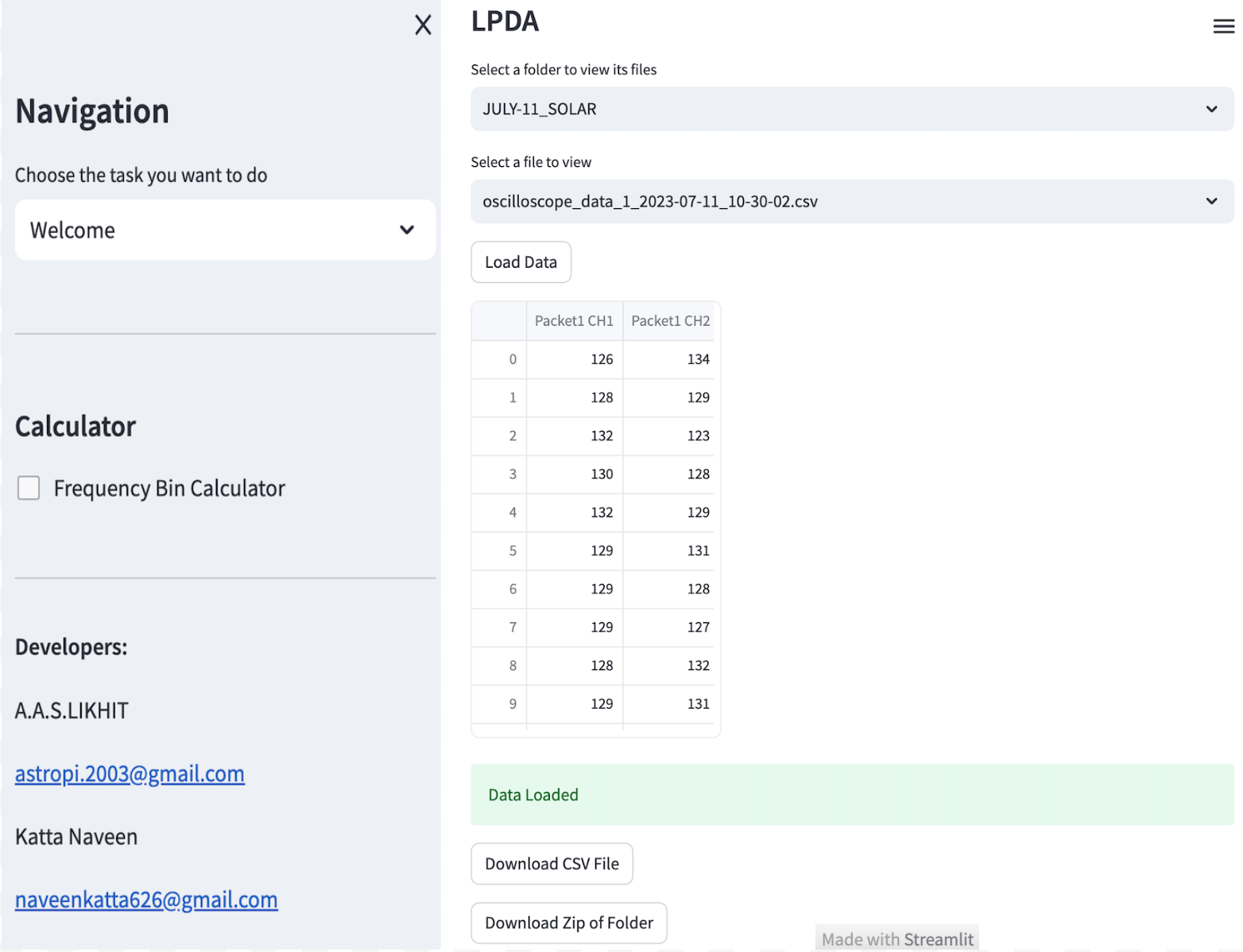}
    \caption{Data Center section of web tool displaying the data of JULY-11 SOLAR observation folder.}
    \label{fig:datacenter}
\end{figure}

\subsection{Data Processing}
This section of the web tool enables users to process observational data. It offers the option to select data from a specific observation for processing, including full data analysis or a quick check by selecting a partial dataset. Users can also choose the FFT segment size for processing. Upon clicking the 'Start Processing' button, the tool begins processing the data and displays results such as initial raw voltage checks, time plots for each channel's raw voltage data, and voltage histograms of an acquisition. Subsequently, it presents the Power Spectrum for each channel and correlated data. It also generates plots of average power across All frequencies with time for Channels 1 and 2, and correlated data, where transit plots can be observed in the absence of RFI.

Additionally, the tool provides Spectrograms for Channels 1 and 2, as well as correlated data. These plots, created using the Plotly library, allow users to zoom into specific regions of the spectrogram, aiding in the identification of RFI-affected bins in the correlated spectrum.

\begin{figure}[h]
    \centering
    \includegraphics[width=3.4in]{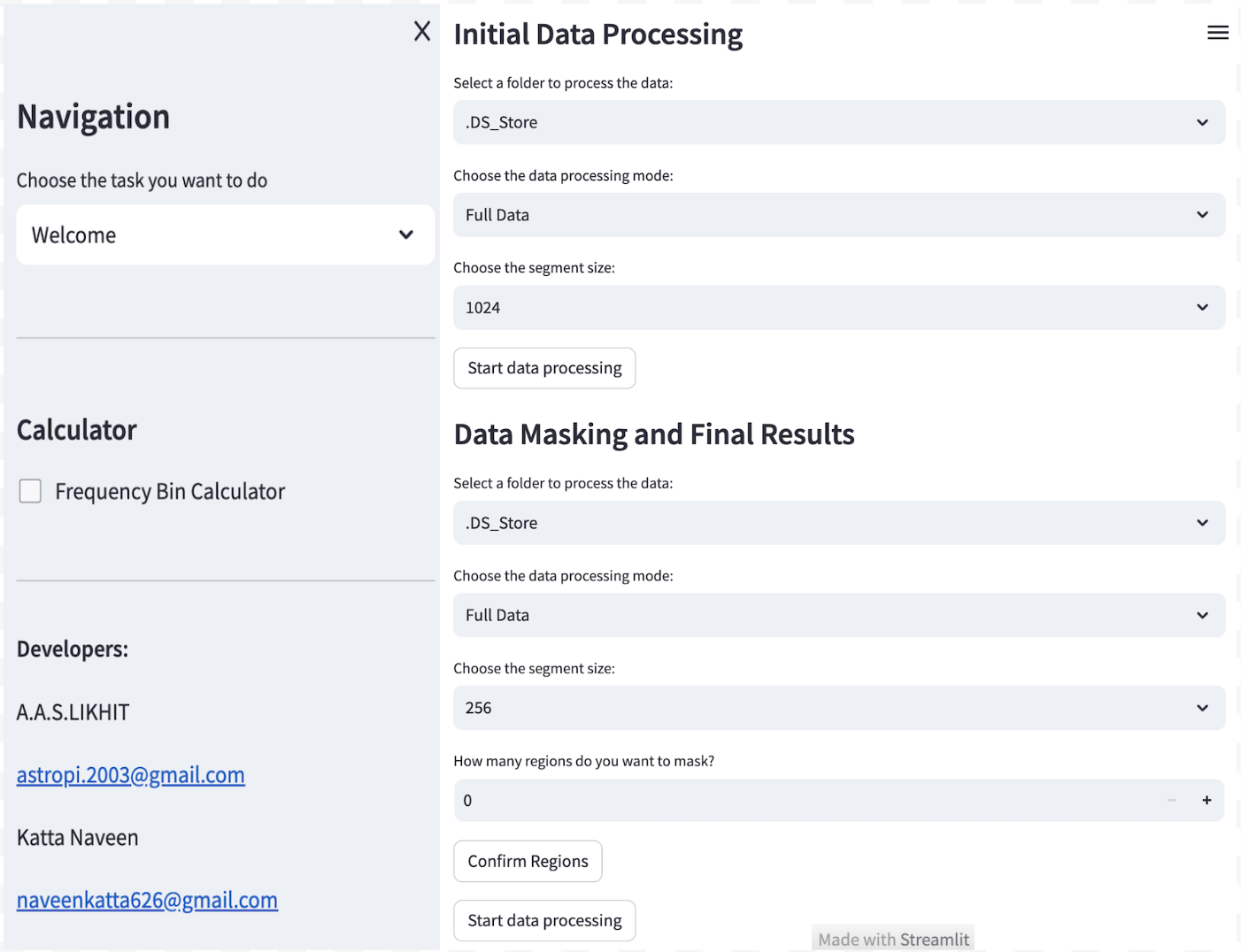}
    \caption{Data Processing section of Web Tool displaying the options for the users input to process the desired observation data.}
    \label{fig:datacenter}
\end{figure}

The advanced data processing feature permits users to specify the number of frequency bin regions to mask and to enter their ranges. After processing with the RFI-affected bins set to zero, the tool produces updated results, including Average Power Across All Frequencies With Time for correlated data, the updated correlated spectrogram, and the fringe pattern at 200MHz. This enables users to view transit plots with RFI regions removed and observe the fringe pattern at 200MHz frequency.

\subsection{Operational Flow}

\begin{figure}[h]
    \centering
    \includegraphics[width=1.8in]{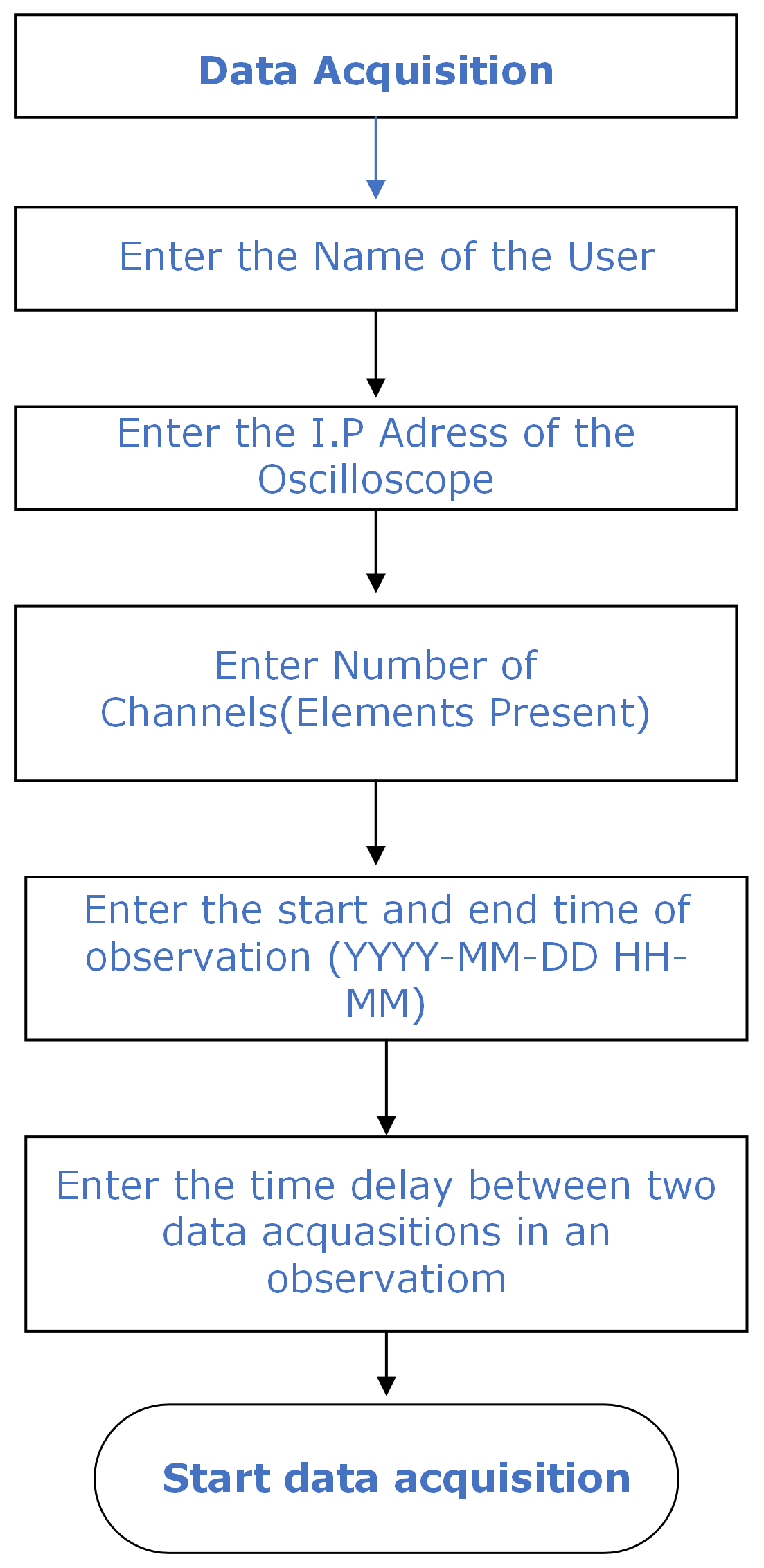}
    \caption{Operational flow of the Data Acquisition Section}
    \label{fig:dawf}
\end{figure}

\begin{figure}[h]
    \centering
    \includegraphics[width=3in]{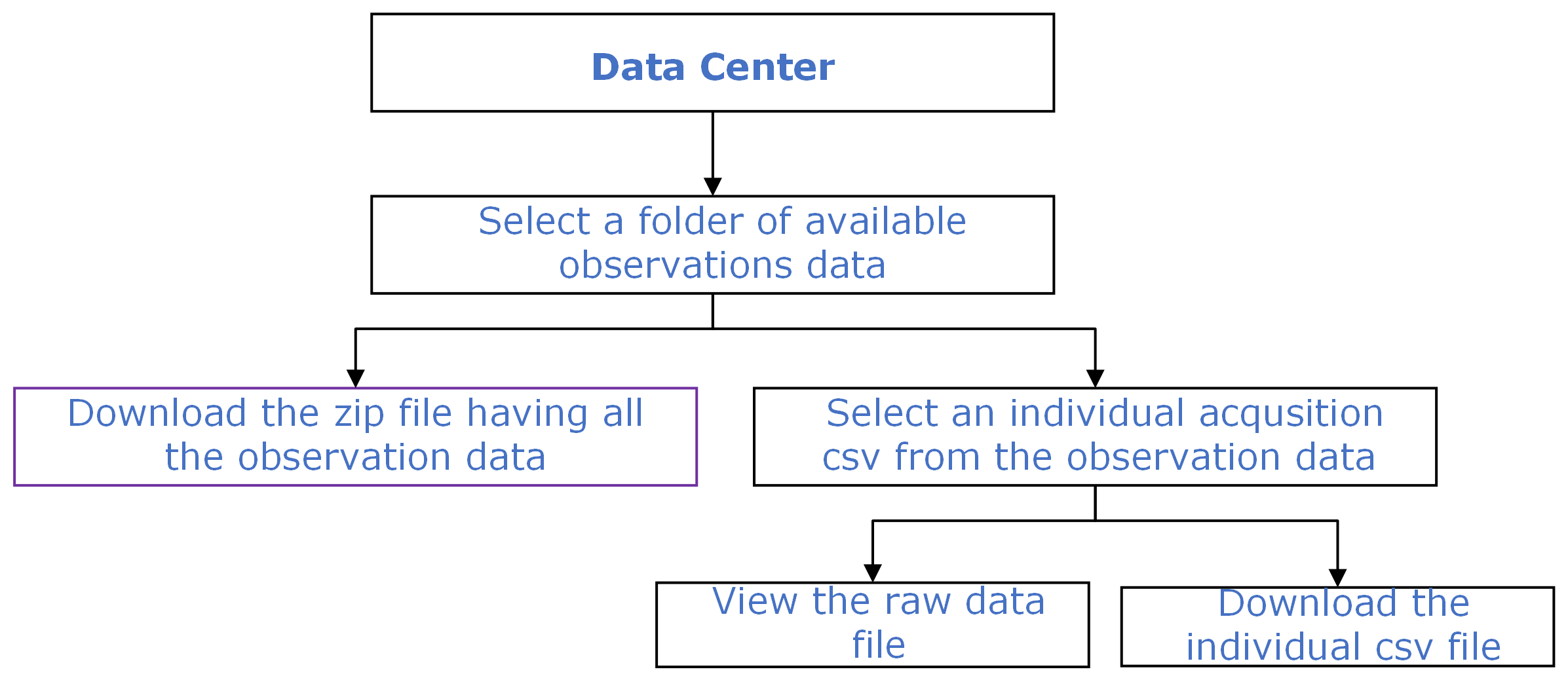}
    \caption{Operational flow of the Data Center Section}
    \label{fig:dcwf}
\end{figure}

\begin{figure}[h]
    \centering
    \includegraphics[width=3.5in]{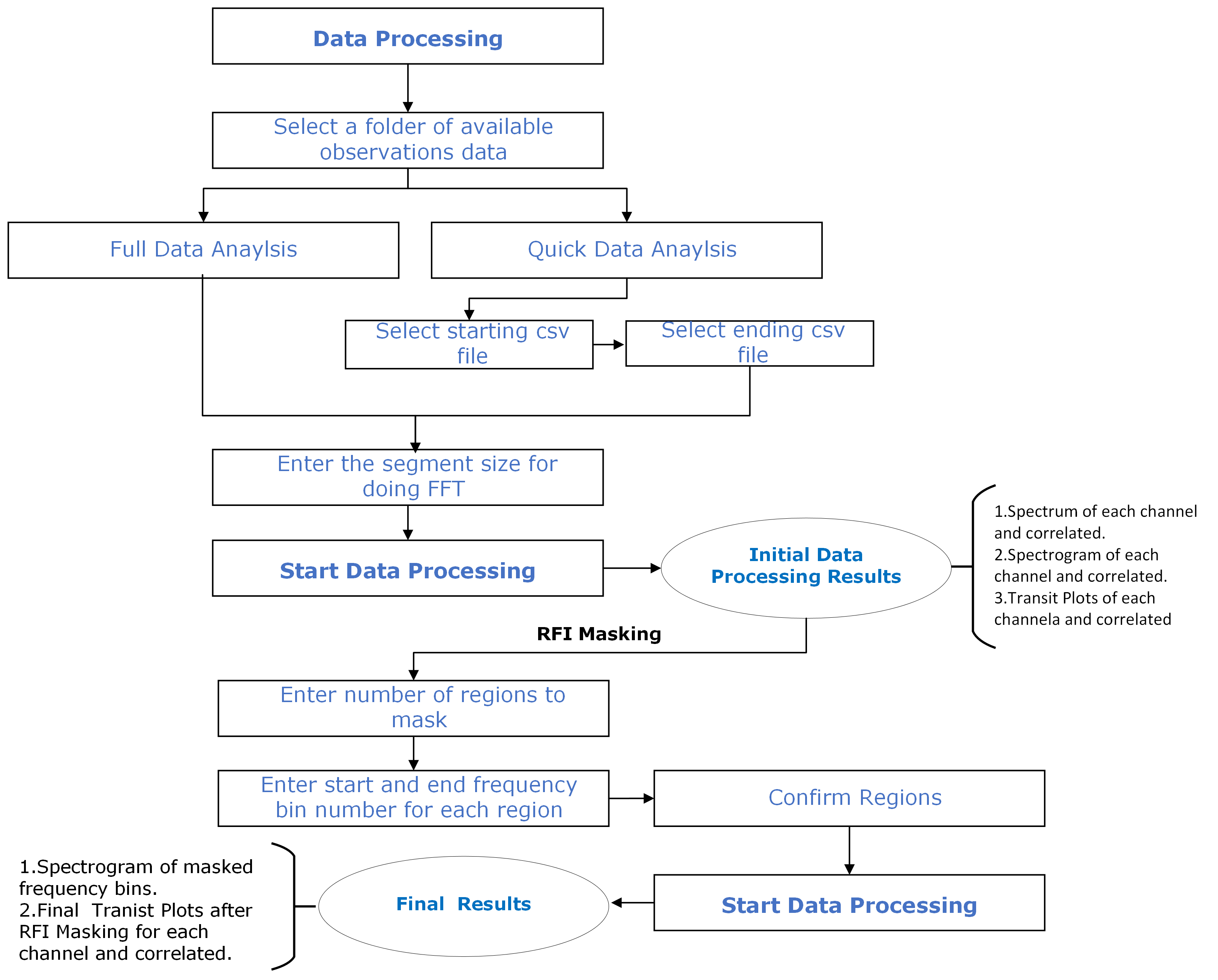}
    \caption{Operational flow of the Data Processing Section}
    \label{fig:dpwf}
\end{figure}

The web tool's functionality is outlined in Figure \ref{fig:functionality}, which depicts its key components: conducting astronomical observations, data processing, and direct data downloading. Detailed workflows for each component are further illustrated in Figures \ref{fig:dawf}, \ref{fig:dcwf}, and \ref{fig:dpwf}, representing the respective operational phases of data acquisition, data center, and data processing.

Figure \ref{fig:dawf} (Data Acquisition) outlines the initial phase, where users establish the settings for capturing data. It starts with the user entering their name and the fixed IP address of the Oscilloscope, ensuring a stable connection for data flow. The user then inputs the number of channels (elements present) and specifies the observation's timeframe using a start and end timestamp (YYYY-MM-DD HH-MM format). Finally, the user sets the interval between data captures within the observation period and name a directory to save the data before initiating data acquisition.

Figure \ref{fig:dcwf} (Data Center) describes the data retrieval and selection process. The user accesses the Data Center, where they select a folder containing available observational data. Options are provided to download the entire dataset as a zip file or to pick an individual CSV file for specific data points. Post-selection, there's a provision to either view the raw data directly within the web tool or to download the individual CSV file for offline use or further analysis.

Figure \ref{fig:dpwf} (Data Processing) describes the comprehensive steps for analyzing the gathered data. It begins with choosing the desired data analysis method—full or quick. For a quick analysis, the user selects a starting and ending CSV file from observation and then proceeds where in full all the data will be taken for the analysis,  enters the segment size for performing Fast Fourier Transform (FFT), and inputs parameters for Radio Frequency Interference (RFI) masking, including the number of regions and their corresponding frequency bins. After confirming these regions, the user can proceed to process the data. The final output includes detailed spectrograms and transit plots, providing insights into the analyzed observations.

\section{Testing and Validation}

During the testing and validation phase, we conducted an observation specifically aimed at capturing a solar transit, and present here results from our observations of 11th July 2023. This event served as an essential test case to evaluate the web tool's capabilities in a real-world scenario. 

\begin{figure}[h]
    \centering
    \includegraphics[width=3.5in]{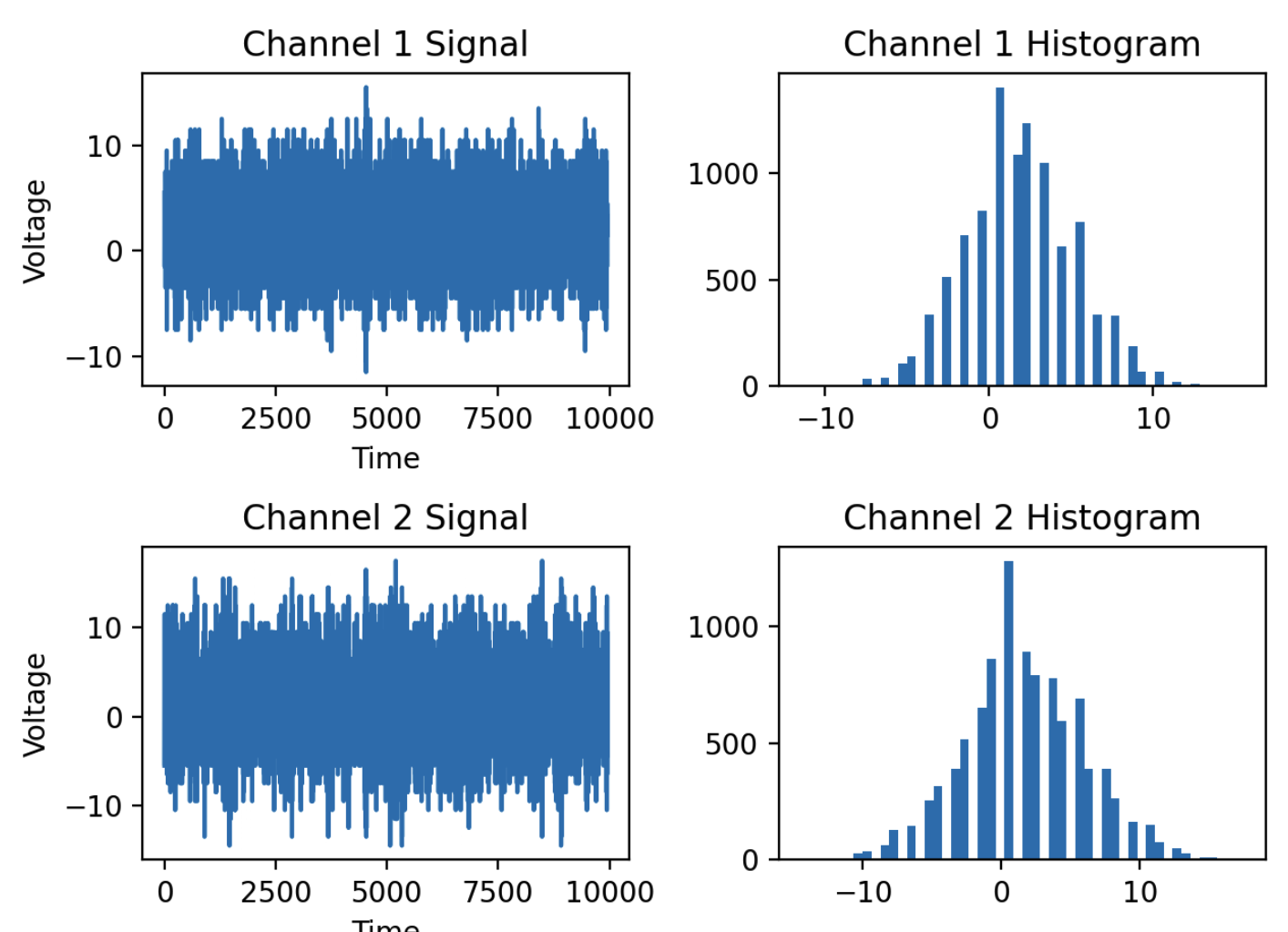}
    \caption{Raw Voltage Monitoring Feature}
    \label{fig:rvmf}
\end{figure}

The data was successfully acquired using the web tool remotely from the Raman Research Institute, located about 85 Km from the observatory, and the data was seamlessly displayed and promptly displayed in the Data Center section of the interface. This demonstrated not only the tool's data handling efficacy but also allowed for immediate dataset downloading.

\begin{figure}[h!]
    \centering
    \includegraphics[width=0.45\textwidth]{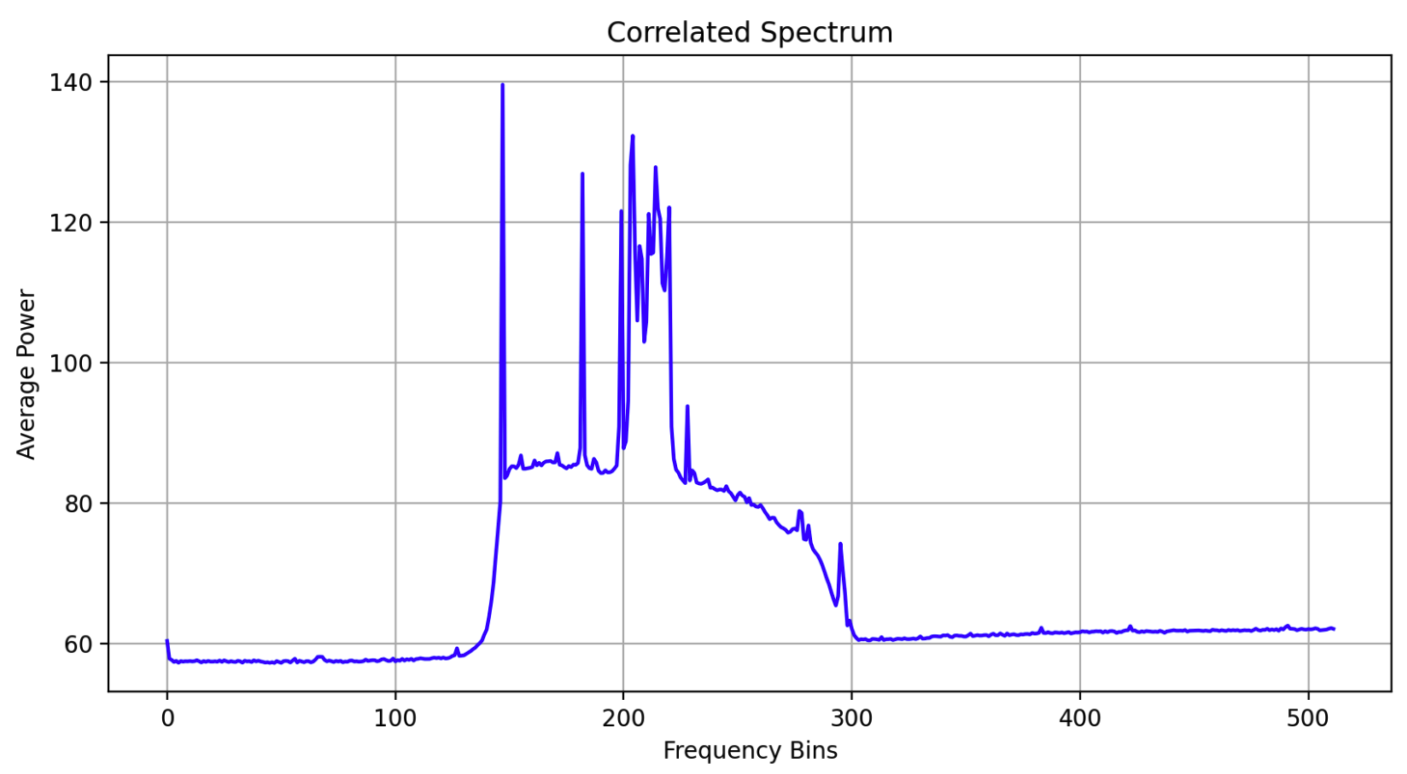}  
        \caption{Power Spectrum}
    \label{fig:ps}
\end{figure}

\begin{figure}[h]
    \centering
    \includegraphics[width=0.45\textwidth]{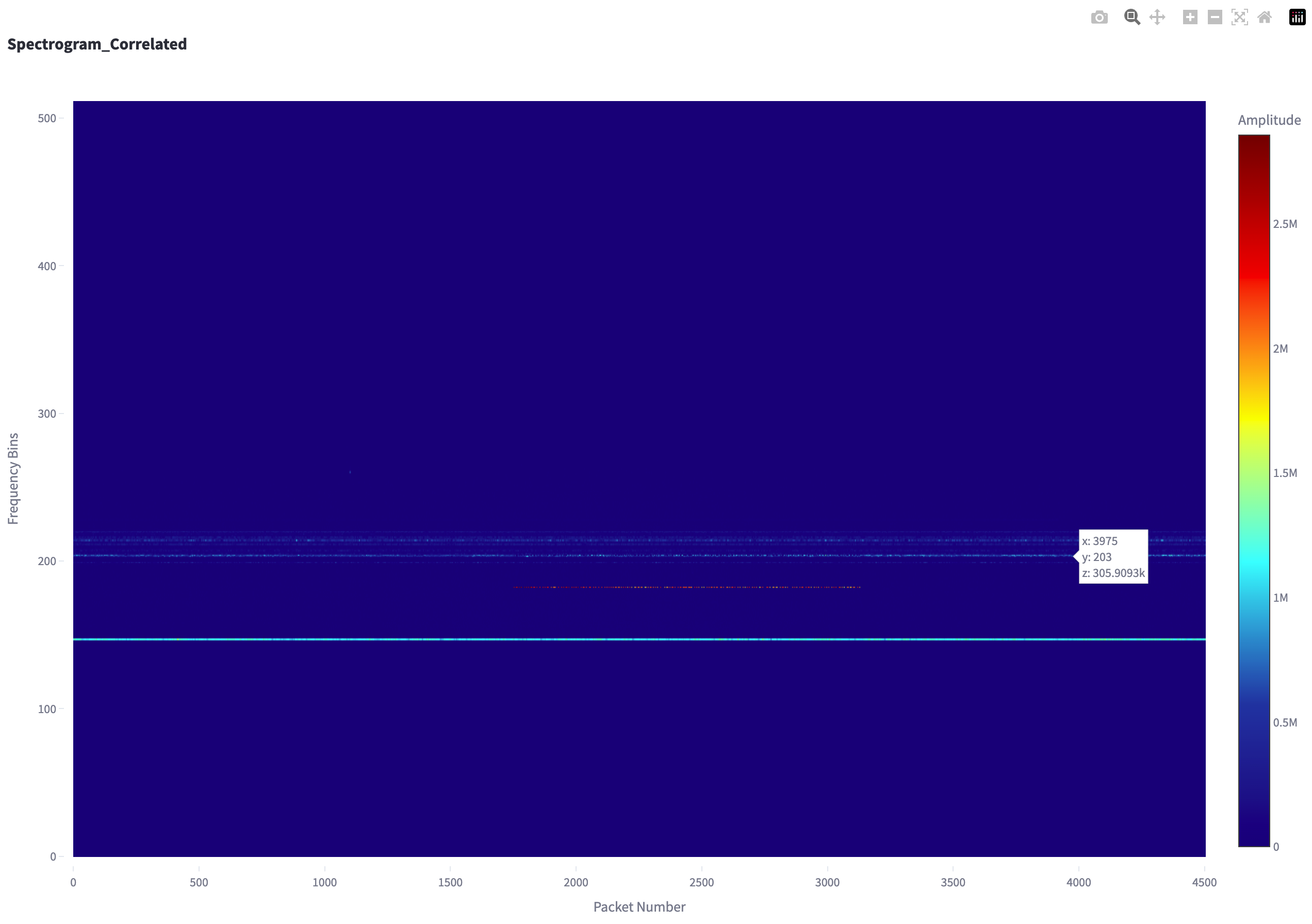}  
    \caption{Spectrogram}
    \label{fig:psg}
\end{figure}

Furthermore, the data was processed directly within the web interface, resulting in successfully generated plots that visualized the solar transit, as shown by the time series and histogram in Figure \ref{fig:rvmf}, and the Power Spectrum and spectrogram in Figures \ref{fig:ps} \(\&\) \ref{fig:psg}.

\begin{figure}[h]
    \centering
    \includegraphics[width=0.45\textwidth]{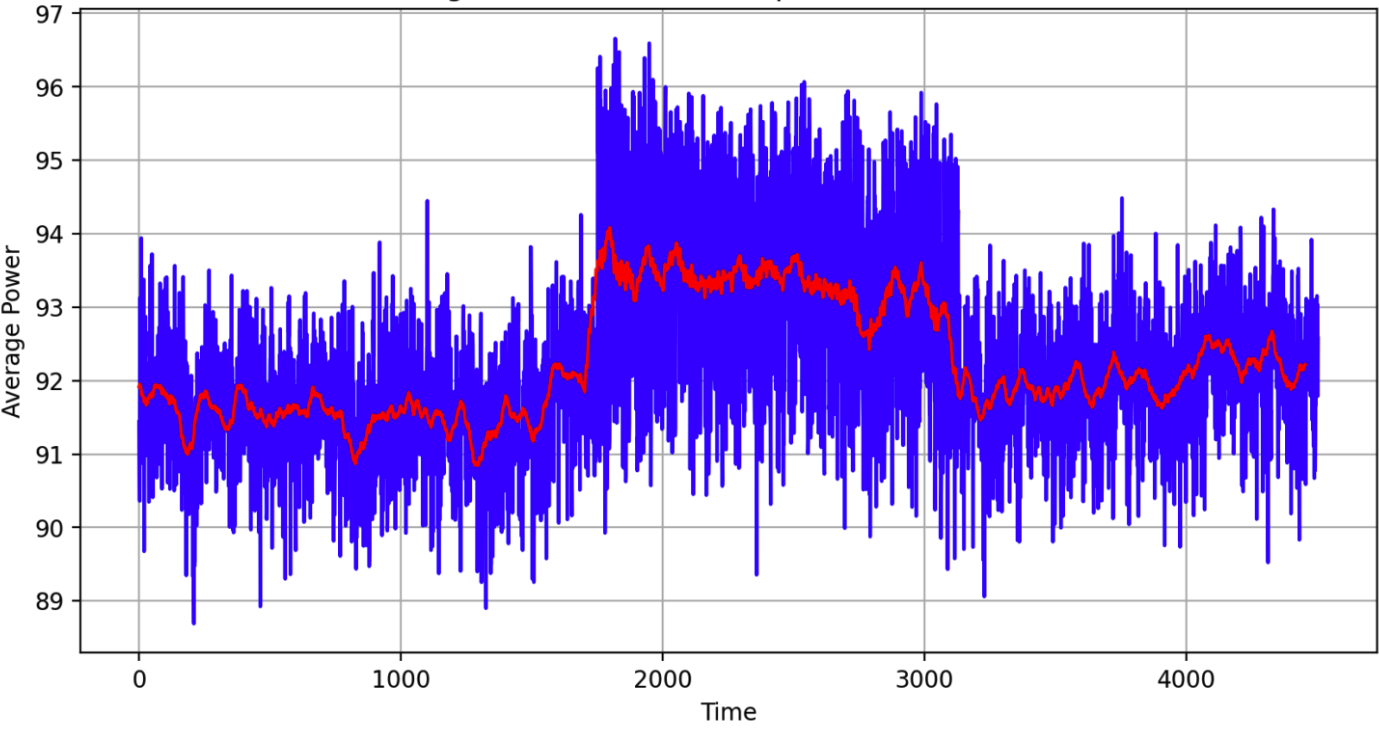}  
        \caption{Average Power Across All Frequencies With Time for correlated data Before RFI Masking}
        \label{fig:tb}
\end{figure}

\begin{figure}[h]
    \centering
    \includegraphics[width=0.45\textwidth]{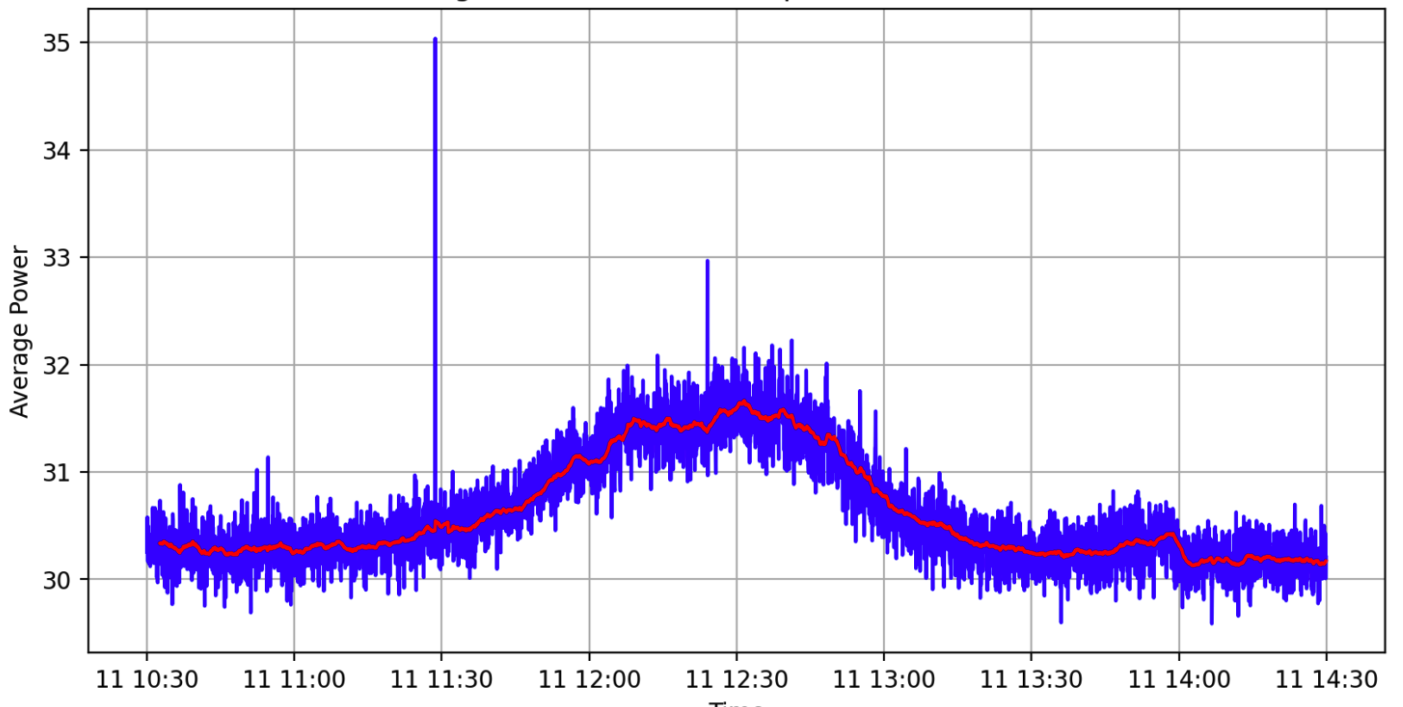}  
        \caption{Average Power Across All Frequencies With Time for correlated data After RFI Masking}
        \label{fig:ta}
\end{figure}

Advanced data processing features were also tested; by specifying frequency bin regions to mask, we obtained a refined plot with minimal RFI interference, which is depicted in Fig. \ref{fig:ta}, with a comparison to the plot before RFI removal in Fig. \ref{fig:tb}. These comprehensive tests confirmed the web tool's functionality for both basic and advanced data analysis tasks.

During the past six months, the tool has successfully operated daily for solar transit and galactic plane transit observations. This extended period of operation has provided substantial evidence of the web tool's reliability and performance, demonstratingAutomated its robustness in consistent astronomical data acquisition and analysis. The consistent results, including the average power across all frequencies with time and the spectrogram analysis, are illustrated in the figures mentioned above.

%We conducted the same analysis for another observation to further validate our findings, and the results are presented below. We monitored the galactic drift and solar transit continuously from July 14th to July 15th. Here we can see \ref{fig:ta_2}, and \ref{fig:ta_2} represents the observed transits of both galactic drift and solar drift in one observation without RFI masking and with masking, respectively. We can see some differences in the plots those are due to the updates that we have pushed into our tool after our first validation.
%\begin{figure}[h]
%    \centering
%    \includegraphics[width=0.45\textwidth]{14_15.png}  
%        \caption{Average Power Across All Frequencies With Time Without Any Masking of the observation of July 14 - July 15 }
%        \label{fig:ta_1}
%\end{figure}
%\begin{figure}
%    \centering
%    \includegraphics[width=1\linewidth]{14_15_1.png}
%    \caption{Average Power Across All Frequencies With Time After Masking}
%    \label{fig:ta_2}
%\end{figure}
\section{Web Tool Adaptability and Future Scope}

The web tool's features and functionalities make it versatile and useful for radio astronomy observations at various remote locations. It integrates seamlessly with the instrumentation by connecting to the IP address of the data acquisition system, such as an oscilloscope. The code is open-source and available on GitHub \citep{GLOT}, allowing users to modify the data processing techniques and even add more elements (antennas) to the array based on their specific observation goals. This flexibility ensures that users from diverse observatories can adapt the tool to their unique setups.

Furthermore, the web tool has the potential to be integrated into larger, more complex systems like the Square Kilometre Array (SKA). The SKA Observatory provides several support tools, available on platforms like GitLab, such as the SKA-OST Sensitivity Calculator and the SKA-OST Simulation for Low Station Behavior \citep{skaost}. These tools are primarily designed for tasks like simulation of results, data processing, and observation management. While they focus on post-observation analysis and sensitivity simulation, our web tool is designed to interact directly with the data acquisition hardware in real-time.

By integrating with SKA’s existing tools, our web tool can complement the SKA infrastructure. For example, SKA tools like the sensitivity calculator simulate and process data after acquisition, but our tool can handle real-time data acquisition, RFI mitigation, and initial processing. This offers observatories the ability to modify the data acquisition process itself and tailor data collection techniques according to the specific requirements of their observations.

Additionally, the web tool gives observatories the freedom to store their data locally on systems or hard disks, removing the need for costly cloud storage solutions like Amazon AWS or Microsoft Azure, which some modern observatories commonly use for data handling. By offering a local storage option, our tool significantly reduces the operational costs associated with data management. Smaller observatories or educational institutions, lacking the budget for large-scale cloud infrastructure, can particularly benefit from this, as it guarantees secure access and processing of data for research purposes.

\subsection{Security Measures}
Currently, the web tool works securely within the observatories using a secure WLAN network. This network ensures that only authorized personnel can access the data acquisition system, minimizing the risk of external or unauthorized access.

We propose linking the web tool to a scheduling system via Google Forms as a future upgrade. This integration would allow users to schedule observations by submitting forms, with access restricted to specific Gmail accounts. We can then use the responses from Google Forms to automatically update the web tool's settings in the Streamlit app, allowing us to schedule and manage observations. This method would not only improve the tool's functionality, but also ensure that only authorized users with permitted Gmail accounts can schedule and access observations, improving the tool's overall security.

\section{Conclusion}

 In this paper, we have presented the design details of an innovative Web tool meant for remote data acquisition and data analysis. This tool is developed for use with the SKA beamforming-related testbed based on the LPDA Array at the Gauribidanur Radio Observatory. This web tool is primarily designed to address the typical needs of a remote observatory. Hence, its usability for the many upcoming small radio astronomy facilities in educational institutions is anticipated. We have presented the salient aspects of this work in this paper. The web tool and ideas presented in this paper can be adapted for the different upcoming facilities. We also foresee using such remote tools for the forthcoming SKA antenna commissioning in the Western Australian deserts. 
 
In conclusion, the web tool we created for the Gauribidanur Radio Observatory significantly addresses the observatory's operational requirements. This tool uses recent developments in data science and technology to offer a scalable and efficient solution for astronomical research. For further details and access to the tool, please refer to the Gauribidanur radio observatory LPDA array Observation Tool (GLOT) repository \citep{GLOT}.

Looking ahead, the utility of systems like the one developed here is not limited to current needs but is expected to grow with upcoming projects such as the Square Kilometre Array (SKA). These tools are essential for modern research and offer a scalable base that can evolve with the field's future requirements. The deployment of our web tool not only addresses the immediate needs of the Gauribidanur Radio Observatory but also supports the ongoing advancement of radio astronomy, enhancing research capabilities and encouraging further progress in the discipline.

%%Appendix

\appendix

\section{Tektronix MSO-3054}
Tektronix MSO-3054 has a bandwidth of 500MHz; the MSO 3054 provides the necessary frequency range for our observations. The oscilloscope features four analogue and sixteen digital channels, allowing for simultaneous data acquisition from multiple sources. A sample rate of 2.5Gs/s on all channels ensures high-resolution sampling of the signals, while the record length of 5 Megapoints on all channels offers substantial memory for recording waveform data. The MSO 3054 also boasts a waveform capture rate of greater than 50,000 waveforms per second, capturing transient events precisely. Furthermore, it is equipped with USB for quick storage, a built-in Ethernet port for network connections, VISA connectivity for remote control, e*Scope for remote viewing and control, and Wave Inspector controls for managing long record lengths with zoom and pan functions, play and pause features, and search and mark capabilities.

%%Use section* for acknowledgements
\section*{Acknowledgements}
We would like to thank the open-source libraries that played an important role in the development of our web tool. We used PyVISA for instrument communication, which allowed for seamless interaction with the data acquisition hardware. Streamlit provided the framework for building the user-friendly web interface, allowing for simple data visualization and remote access. Finally, we utilized Selenium for web automation, which facilitated efficient interaction with the oscilloscope's web interface.
\vspace{-1em}
 
\bibliography{bibliography}
\end{document}